\documentclass{aa}  

\usepackage{graphicx}
\usepackage{txfonts}
\usepackage{hyperref}
\usepackage{xcolor}
\usepackage[normalem]{ulem}

\begin{document}

   \title{MUSE crowded field 3D spectroscopy in NGC\,300}
%   \title{Spectroscopy of bright BA-type stars in NGC\,300 with MUSE}

   \subtitle{II. Quantitative spectroscopy of BA-type
   supergiants %and a\\ first detection of post-AGB stars beyond the Local Group
   %\thanks{Based on observations collected at the European
   %Organisation for Astronomical Research in the Southern Hemisphere
   %under ESO programme 094.D-0116(A).}
   }

\author{G. Gonz\'alez-Tor\`a
          \inst{1,2,3}
          \and
          M. A. Urbaneja\inst{3}
          \and
          N. Przybilla\inst{3}
          \and
          S. Dreizler\inst{4}
          \and
          M. M. Roth\inst{5}
          \and
          S. Kamann\inst{2}
          \and
         N. Castro\inst{5}
          }

   \institute{European Southern Observatory (ESO),
   Karl-Schwarzschild-Str. 2, 85748 Garching bei München, Germany\\
        \email{Gemma.GonzaleziTora@eso.org}
   \and
        Astrophysics Research Institute, Liverpool John Moores
        University, 146 Brownlow Hill, Liverpool L3 5RF, United Kingdom\ 
    \and
            Institut f\"ur Astro- und Teilchenphysik, Universit\"at Innsbruck, Technikerstr. 25/8, 6020 Innsbruck, Austria \
             \and
             Institute for Astrophysics, University of G\"ottingen, Friedrich-Hund-Platz 1, 37077 G\"ottingen, Germany \
             \and
            Leibniz-Institut f\"ur Astrophysik (AIP), An der Sternwarte 16, 14482 Postdam, Germany}

   \date{Received \today; accepted -}

% \abstract{}{}{}{}{} 
% 5 {} token are mandatory
 
  \abstract
   {}
   {A quantitative spectral analysis of BA-type supergiants and bright giants 
   in an inner spiral arm region of the nearby spiral galaxy NGC\,300 
   is presented, based on observations with the 
   Multi Unit Spectroscopic Explorer (MUSE) on the European Southern Obsevatory, Very Large Telescope (ESO, VLT). 
   The flux-weighted gravity--luminosity relationship (FGLR), a stellar
   spectroscopic distance determination method for galaxies,
   is extended towards stars at lower luminosities.}
   {Point spread function fitting 3D spectroscopy was performed with
   PampelMUSE on the datacube. 
   The 16 stars with the highest signal-to-noise ratios ($S/Ns$)  are classified with
   regard to their spectral type and luminosity class using Galactic
   templates. They were analysed using hybrid non-local thermodynamic equilibrium (non-LTE) model spectra to fit 
   the strongest observed hydrogen, helium, and metal lines in the
   intermediate-resolution spectra. Supplemented by
   photometric data, this facilitates fundamental stellar
   parameters and interstellar reddening which have yet to be determined.}
   {Effective temperatures, surface gravities, reddening $E(B-V)$,
    bolometric magnitudes and luminosities, as well as radii and masses are presented
    for the sample stars. The majority of the objects follow the
    FGLR as established from more luminous BA-type supergiants in
    NGC\,300. An increase in the scatter in the flux-weighted
    gravity--luminosity plane is observed at these lower luminosities,
    which is in line with predictions from population synthesis models.
    %For four object much lower flux-weighted gravities are found than
    %predicted by the FGLR at the given luminosities. Moreover, unlike
    %the other sample stars, these objects show H$\alpha$ in emission.
    %The observational evidence points to a post-AGB nature of these
    %stars, the first such findings in a galaxy beyond the Local Group.
    }
    {}

   \keywords{Galaxies: individual (NGC\,300) -- Galaxies: distances and
   redshifts -- Stars: early-type --  Stars: fundamental parameters -- 
   supergiants}

   \maketitle
%
%-------------------------------------------------------------------

\section{Introduction}
Historic turning points in astronomy were the identification
of the true nature of galaxies by resolving their stellar content
\citep{1929ApJ....69..103H} and the identification of stellar populations \citep{Baade44}, which facilitated an understanding of galaxy evolution which has yet to be developed. Currently, this can be achieved for
galaxies out to distances in the nearest clusters of galaxies using the Hubble Space
Telescope or the ground-based 8-10\,m telescopes.

Far more challenging is the spectroscopy of individual stars in other
galaxies. Early spectroscopic observations with 4\,m-class telescopes
showed that the visually brightest stars in the Local Group galaxies
and beyond are massive blue supergiants of late B and A spectral types
(BA-type supergiants), reaching
absolute visual magnitudes of $M_V$\,$\simeq$\,$-9.5$
\citep[e.g.][]{1979ApJ...232..409H,1987AJ.....94.1156H}. These stars are
typically evolved objects crossing the Hertzsprung-Russell diagram (HRD) for
the first time towards the red after terminating their main-sequence
phase as OB-type stars. They reach their high visual brightness because
of the low bolometric corrections at these temperatures, that is~$M_V$\,$\simeq$\,$M_\mathrm{bol}$.

The potential of the stars for quantitative analyses over larger
distances was previously identified, leading to the first
determination of atmospheric parameters and chemical abundances of the
Galactic prototype A-type supergiant \object{Deneb} by \citet{1961ZA.....51..231G}
and the brightest BA-type supergiants in the Magellanic Clouds
\citep{1968MNRAS.139..313P,1971MNRAS.152..197P,1972MNRAS.159..155P,wolf2,wolf3},
all based on Coud\'e spectra recorded on photographic plates.
However, systematic studies of samples of A-type supergiants in the
Milky Way and the Small Magellanic Cloud (\object{SMC}) at high
spectral resolution were only performed much later
\citep{1995ApJS...99..659V,1995ApJ...449..839V,1999ApJ...518..405V},
based on Coud\'e and Echelle spectra recorded with CCD detectors.
At about the same time, the first quantitative analyses of a few individual 
BA-type supergiants in galaxies of the Local Group beyond the Magellanic Clouds 
were undertaken based on Echelle spectra taken with Keck and the
European Southern Observatory (ESO) Very
Large Telescope (VLT) in
\object{M33}, \object{M31}, \object{NGC6822,} and \object{WLM}
\citep{1995ApJ...455L.135M,2000ApJ...541..610V,2001ApJ...547..765V,
2003AJ....126.1326V}.

A more efficient use of telescope time was promised by multi-slit
spectrographs at intermediate spectral resolution by increasing the
number of simultaneously observed targets to about 20, for example with the visual and 
near UV FOcal Reducer and low dispersion Spectrograph (FORS)
on the VLT \citep{1998Msngr..94....1A} or the Low Resolution Imaging Spectrometer (LRIS)
on Keck
\citep{1995PASP..107..375O}. BA-type
supergiants are bright enough to be observable in galaxies out to the Virgo cluster 
with 8-10m telescopes \citep{1995kudritzki}. Practical limitations, 
in particular crowding at seeing-limited conditions, restrict quantitative studies 
to about a 7\,Mpc distance, such as in \object{NGC3621} 
\citep{2001ApJ...548L.159B,2014ApJ...788...56K} or the maser-host galaxy \object{NGC4258}
\citep{2013ApJ...779L..20K}.
Studies encompassed nearby metal-poor irregular galaxies such as WLM 
\citep{2008ApJ...684..118U}, \object{IC1613} \citep{2018ApJ...860..130B},
or \object{NGC3109} \citep{2014ApJ...785..151H}. 
At the focus of investigations stood galactic abundance gradients
\citep[e.g.][]{Kudritzki08} and the comparison of stellar and nebular
abundance indicators
\citep[e.g.][]{2005ApJ...622..862U,2009ApJ...700..309B,2016ApJ...830...64B}.
A closely related field was the investigation of 
the galaxy mass--metallicity relationship
\citep[e.g.][]{1979A&A....80..155L,2004ApJ...613..898T}
based on stellar indicators \citep[][]{2012ApJ...747...15K}, which
help to overcome uncertainties from employing strong-line-methods
for nebular analyses \citep{2008ApJ...681.1183K,2013ApJ...765..140A}. 
In addition, contributions were
made to establish a novel spectroscopic distance indicator, the flux-weighted gravity-luminosity relationship
\citep[FGLR,][]{2003ApJ...582L..83K,Kudritzki08} by
concentrating on important calibrators of the extragalactic
distance scale such as the Large Magellanic Cloud 
\citep[\object{LMC},][]{2017AJ....154..102U} and \object{M33} \citep{2009ApJ...704.1120U}.

A particular role for stellar studies in systems of the nearby
galaxy groups play those towards the filamentary Sculptor Group
\citep{2003A&A...404...93K} as they are located 
close to the Galactic southern pole, that is these are the galaxies affected least by
Galactic extinction. In the foreground -- at distances shortly below
2\,Mpc -- are \object{NGC55} \citep{2012A&A...542A..79C,2016ApJ...829...70K} 
and \object{NGC300}. As a spiral galaxy seen nearly face-on, the latter one
has attracted many quantitative investigations using multi-object slit spectroscopy 
\citep{2002ApJ...567..277B,2004ApJ...600..182B,2009ApJ...700..309B,2003ApJ...584L..73U,
2005ApJ...622..862U,Kudritzki08}

Innovation on the instrumental side in the form of integral field
spectroscopy (IFS) or 3D spectroscopy promises to increase the multiplex 
further by more than an order of magnitude. This technique allows for a spectrum for 
each pixel across an image to be obtained
simultaneously. It is an extremely powerful and fast tool to perform 
spectroscopy on extended objects. Groundbreaking in this context is the Multi
Unit Spectroscopic Explorer \citep[MUSE,][]{2014Msngr.157...13B} on
the ESO VLT, which combines a wide field of view with high spatial
sampling (1\arcmin$\times$1\arcmin~field with 0\farcs2 sampling).
The first surveys of the massive star populations in nearby galaxies have already 
started with MUSE. Demonstrations were made for dense star clusters
in the LMC \citep{2018A&A...614A.147C,2021A&A...648A..65C} and SMC
\citep{2020A&A...634A..51B}, and for wider fields in the Sculptor
Group galaxies NGC\,300
\citep[][henceforth Paper~I]{2018A&A...618A...3R} and
\object{NGC7793} \citep{2020MNRAS.493.2410W}. 
%Still missing is the
%demonstration of the usefulness of MUSE data for detailed
%quantitative spectral analyses of massive stars. Complications in this
%context arise because the blue spectral range, which is traditionally
%employed for massive star analyses, is not covered by MUSE.
%\citep[a motivation in proposing the blue-sensitive BlueMUSE,][]{2019arXiv190601657R}. 

The present study aims to provide a detailed analysis of MUSE
spectra of BA-type supergiants and bright giants in NGC\,300 in one field 
from Paper~I based on synthetic spectra that account for deviations
from local thermodynamic equilibrium (LTE). The paper is organised as follows: 
a brief overview of the observations, the data reduction, and the
source extraction is given in Section~\ref{sec:obs}. Section~\ref{sec:analysis} 
gives an overview of the employed models as well as
describes the spectral analysis and the resulting atmospheric and
fundamental stellar parameters. A discussion of the results, in
particular focussing on an extension of the FGLR towards lower luminosities, 
is presented in Section~\ref{sec:discussion}. Finally, prospects for future work are 
discussed in Section~\ref{sec:conclusion}. 

\begin{figure}[t]
\centering
\includegraphics[width=1.\linewidth]{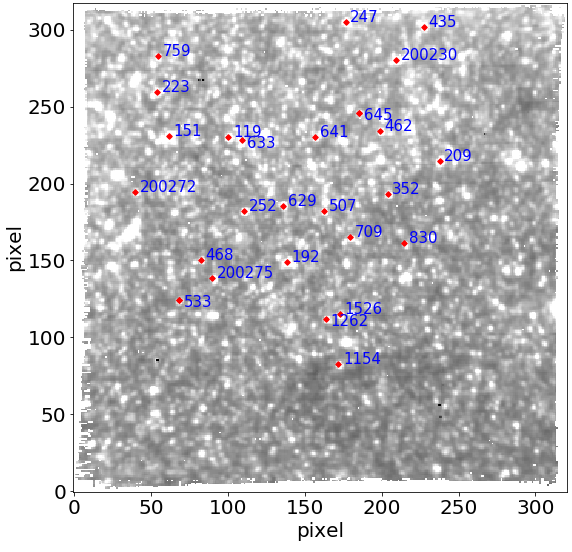}
\caption{Chart of NGC\,300 field (i) with the programme stars 
marked in red and their ID numbers in blue (according to Paper~I).
The image is stacked over all recorded wavelengths in the datacube.}
\label{fig:photometry}
\end{figure}

%--------------------------------------------------------------------

\section{Observations and data reduction}\label{sec:obs}

The spectroscopic data were obtained using MUSE
\citep{2014Msngr.157...13B}, which is located on the Unit Telescope 4 (UT4) 
Nasmyth focus of the VLT at Paranal Observatory in Chile. 
The wide field mode (WFM) with 1\arcmin$\times$1\arcmin~spatial coverage 
and 0\farcs2 sampling was used. The data discussed here were taken prior 
to the adaptive optics (AO) improvement as implemented via the GALACSI
module to the MUSE facility.
 
Therefore, the pointing observed under the best seeing conditions
(FWHM\,=\,0\farcs47-0\farcs59, measured from the data) was investigated, that is 
the one that promises the best spatial resolution and highest
signal-to-noise ratio ($S/N$) per extracted spectrum. This corresponds to
field (i) of Paper~I (see their Fig.~1).
The three observations of this field were carried out 
on October 30, 2014 and on November 24 and 25, 2014, with an exposure 
time of 1800\,s each at an airmass of 1.03-1.05.
The data obtained with the WFM have an intermediate spectral resolving
power of $R$\,=\,$\lambda/\Delta\lambda\,\approx\,$1800-3600, 
a dispersion of 1.25\,{\AA}\,pixel$^{-1}$, and cover the wavelength range 
from 4650 to 9300\,{\AA} (extended wavelength range). 

The initial reduction was achieved with the MUSE pipeline V1.0
\citep[see Paper~I for details]{2020A&A...641A..28W}.  %Peter said it was too much detail---
%master bias and master flat-field, 3 lamps (HgCd, Xe, Ne) were used to derive the 
%wavelength solution. In addition, 11 sky flat-fields 
%were combined to create a 3D illumination correction. Finally, the instrument 
%artefacts were removed using a bad-pixel table.
The final data were produced in the form of a datacube, 
and the spectra of 606 individual sources were extracted using 
the PampelMUSE software \citep{kamann13}.

Out of these 606 extracted sources, 26 were classified as
late-B to early-A supergiants or bright giants in Paper~I with a minimum S/N$>$7, which was imposed in order for our analysis to provide meaningful results. They are identified in 
Fig.~\ref{fig:photometry}, which shows NGC\,300 field (i) in spaxel coordinates.
The stars cover deprojected galactocentric distances of 1.4 to 1.9\,kpc when adopting the following NGC\,300 galaxy parameters from the HyperLEDA database \citep{2014A&A...570A..13M}: 
galaxy centre RA(J2000)\,=\,00:54:53.54/DEC(J2000)\,=\,$-$37:41:04.3, an inclination between the line of sight and the polar axis of the galaxy 48\fdg5, a major axis position angle of 113\fdg2,
and a distance of 1.86\,Mpc \citep{Rizzi2006}.

\begin{table*}
\caption{Observational data for field (i) stars identified as BA-type (super)giants in Paper~I. ID numbers are taken from Paper~I.}
        \label{tab:photometry}
        \small
        \centering
        \begin{tabular}{l l l r r r l} 
                \hline \hline
ID     &    RA(J2000)  &     DEC(J2000) &   $V$            &          $(B-V)$ & spec.~$S/N$ & Comment \\  
\hline                                                                                                        
119    &        00 54 44.0 &    $-$37 41 50.9 &   20.49$\pm$0.04 &    0.10$\pm$0.03 &  20.3 &\\  
151    &    00 54 43.4 &        $-$37 41 51.1 &   20.84$\pm$0.04 &    0.04$\pm$0.03 &  23.3 &\\  
192    &        00 54 42.7 &    $-$37 42 07.3 &   20.88$\pm$0.04 &    0.04$\pm$0.03 &  19.4 &\\  
209    &    00 54 41.0 &    $-$37 41 54.3 &   20.91$\pm$0.04 &    0.15$\pm$0.03 &  18.5 &\\  
223    &    00 54 42.1 &    $-$37 41 36.1 &   21.06$\pm$0.04 & $-$0.03$\pm$0.03 &  14.7 & \\  
247    &    00 54 43.2 &    $-$37 42 00.7 &   21.10$\pm$0.04 &    0.02$\pm$0.03 &  17.8 &\\
252    &    00 54 41.5 &        $-$37 41 41.1 &   21.03$\pm$0.04 &    0.07$\pm$0.03 &  17.5 & beyond model grid coverage \\    
352    &    00 54 41.2 &        $-$37 41 36.8 &   21.30$\pm$0.04 & $-$0.02$\pm$0.03 &  14.9 & in H\,{\sc ii} region \\  
435    &    00 54 41.6 &    $-$37 41 58.6 &   21.67$\pm$0.04 &    0.28$\pm$0.03 &  15.4 & in H\,{\sc ii} region \\  
462    &    00 54 44.1 &    $-$37 41 45.2 &   21.61$\pm$0.04 &    0.51$\pm$0.03 &  13.4 &\\  
468    &    00 54 41.7 &    $-$37 41 50.4       &   21.58$\pm$0.04 &    0.07$\pm$0.03 &  13.2 &\\  
507    &    00 54 43.7 &    $-$37 42 07.0 &   21.63$\pm$0.04 &    0.05$\pm$0.03 &  11.8 &\\ 
533    &    00 54 42.3 &    $-$37 42 00.8 &   21.78$\pm$0.04 &    0.40$\pm$0.03 &   8.7 & beyond model grid coverage \\
629    &    00 54 44.4 &    $-$37 41 58.2 &   21.86$\pm$0.04 &    0.08$\pm$0.03 &  11.1 &\\  
633    &    00 54 42.8 &    $-$37 42 00.1 &   21.72$\pm$0.04 &    0.16$\pm$0.03 &  10.2 &\\  
641    &    00 54 40.8 &    $-$37 41 58.7       &   21.77$\pm$0.04 &    0.12$\pm$0.03 &  10.3 &\\  
645    &    00 54 42.0 &    $-$37 42 04.1 &   21.64$\pm$0.04 &    0.16$\pm$0.03 &  9.6 &\\  
709    &    00 54 42.4 &    $-$37 41 51.1 &   21.71$\pm$0.04 &    0.07$\pm$0.03 &  10.4 &\\  
759    &    00 54 43.2 &    $-$37 41 51.5 &   21.83$\pm$0.04 &    0.16$\pm$0.03 &   7.2 &\\ 
830    &    00 54 43.9 &    $-$37 42 12.2 &   ...            &    ...           &   8.2 &  no Johnson photometry \\
1154   &    00 54 41.9 &    $-$37 41 48.0 &   ...            &    ...           &   7.2 &  no Johnson photometry \\
1262   &    00 54 44.1 &    $-$37 41 40.5 &   22.00$\pm$0.04 &    0.47$\pm$0.03 &   8.6 &  beyond model grid coverage \\
1526   &    00 54 41.4 &    $-$37 42 04.9 &   22.30$\pm$0.04 &    0.27$\pm$0.03 &   7.7 &  beyond model grid coverage \\
200230 &    00 54 42.1 &    $-$37 42 20.4       &   21.56$\pm$0.04 &    0.42$\pm$0.03 &  16.2 &\\  
200272 &    00 54 42.3 &    $-$37 42 14.5       &   21.62$\pm$0.04 & $-$0.01$\pm$0.03 &  11.7 & in H\,{\sc ii} region \\  
200275 &    00 54 42.1 &    $-$37 42 14.0 &   21.47$\pm$0.04 &    0.08$\pm$0.03 &  13.2 & in H\,{\sc ii} region \\  
\hline
        \end{tabular}
\end{table*} 

The spectroscopic data were complemented by Johnson photometry in the $B$ and $V$ bands obtained with the Wide Field Imager (WFI) at the ESO/MPI 2.2\,m telescope on La
Silla. These data were previously employed in stellar studies of NGC\,300
\citep{2002ApJ...567..277B,2004ApJ...600..182B}. Details of the
calibration and reduction of these data were discussed by
\citet{2002AJ....123..789P}. 
%Peter comment: point out that ESO WFI has much lower spatial resolution

The observational properties of the 26 stars are summarised in Table~\ref{tab:photometry}. The stars were identified by an ID number (adopted from Paper~I), the RA and DEC coordinates are given, Johnson $V$ magnitudes and $(B-V)$ colours are stated, the $S/N$  of the extracted spectra are indicated, and comments on individual stars are given, where appropriate. For two stars, no Johnson photometric data are available. Four objects show, in their spectra, emission lines that are characteristic of low-excitation H\,{\sc ii} regions, such as [N\,{\sc ii}] and [S\,{\sc ii}], and very weak [O\,{\sc iii}]. The stellar hydrogen lines are therefore expected to be contaminated by nebular emission. Moreover, the stars are too cool to excite the nebulae; therefore, it is likely that nearby OB-type stars contribute to the recorded spectra. Finally, four more stars will later turn out to have atmospheric parameters that lie beyond those covered by our model grids. Consequently, only 16 objects fulfilled the criteria for a quantitative analysis, which is discussed in the following.

%-----------------------------------------------------------------
\section{Analysis}\label{sec:analysis}

\subsection{Models}
We adopt the modelling methodology developed by \citet{Norbert06} for the analysis of our final sample 
of 16 stars. Very briefly, a combination of model atmosphere structures calculated under the assumption 
of local thermodynamic equilibrium (LTE) and a detailed non-LTE level population as well as line-formation 
calculations are employed. The reader is referred to \citet{Norbert06} for an extensive discussion 
of the advantages and drawbacks of this hybrid approach, as well as its applicability limits. To ensure 
these limits are respected, the analyses are restricted to the
brightest late B-type and A-type giant and 
supergiant stars in the observed sample, which initially corresponds to the 26 stars shown 
in Table~\ref{tab:photometry}. For the reasons already mentioned, we obtain a final sample of 
16 stars. %Furthermore, a minimum S/N$>$7 is imposed, in order for the analysis to provide meaningful results.
%\textbf{We want to note that unlike in previous applications of the
%models, MUSE does not cover the blue spectral region of the target.
%Among the Balmer lines only H$\alpha$ and H$\beta$ are observed, which
%may be influenced in supergiant stars by the presence of a sufficiently strong stellar wind
%\citep[e.g.][]{McCarthyetal97}. However, the target stars are of
%luminosity class Ib and even II, which are characterised by symmetric
%line profiles throughout all members of the Balmer }

\begin{figure}[t]
\centering
\includegraphics[width=\linewidth]{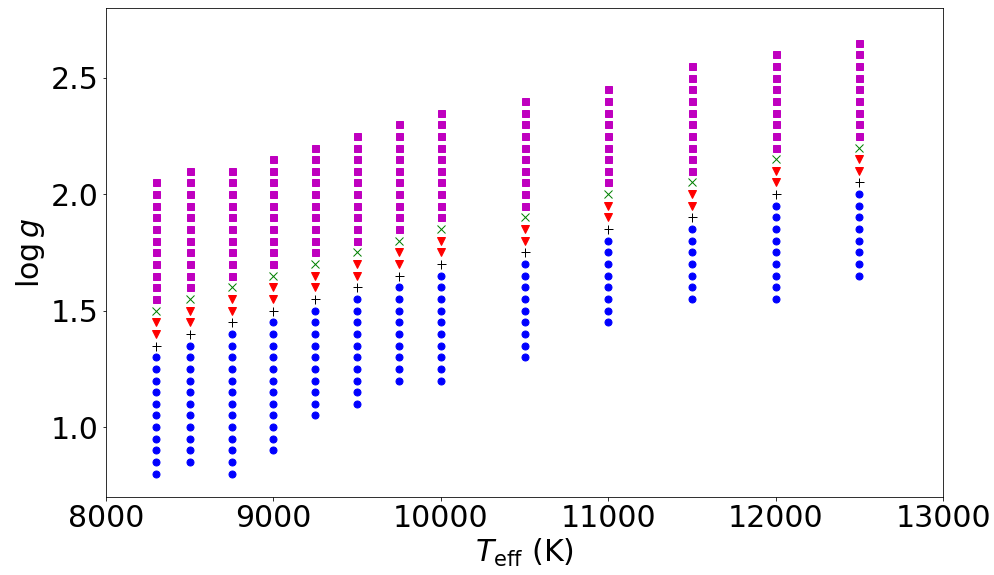}
\caption{Parameter coverage of the model atmosphere grid used in the
present work. The different symbols refer to the microturbulent
velocities $\xi$ adopted in the computation of the models. Blue circles indicate $\xi=8$\,km\,s$^{-1}$, black pluses 7\,km\,s$^{-1}$, red triangles 6\,km\,s$^{-1}$, green crosses 5\,km\,s$^{-1}$, and magenta diamonds 4\,km\,s$^{-1}$.}
\label{fig:models}
\end{figure}

Following previous work on the quantitative analysis of low resolution optical spectra of extragalactic BA-type supergiants (see for example \citealt{Kudritzki08}), we initially considered a grid of model atmospheres exploring the parameter space defined by the effective temperature  $T_{\mathrm{eff}}$, surface gravity $\log g$, global metallicity $\left[Z\right]$, and microturbulence  $\xi$. 
The location of the stars in the disk of the galaxy at
galactocentric distances between 1.4 and 1.9\,kpc, 
in combination with the NGC\,300 metallicity gradient \citep{Kudritzki08,2015ApJ...805..182G}, 
constrains the metallicity at an average $[Z]$\,=\,$-$0.2 dex, that is to say slightly lower than solar. 
To explore the limitations of our models, we used the available grids with both $[Z]$\,=\,0\,dex and 
$[Z]$\,=\,$-$0.15\,dex 
and found that the results of the analysis using both $[Z]$ changed by
$\lesssim 200$\,K in temperature, and $\lesssim 0.05$\,dex in $\log$ gravity, 
so both results agree within the error limits.
Furthermore, following the results obtained in the high resolution work by \cite{Przybilla02}, we adopted a dependency of the microturbulence with $T_{\mathrm{eff}}$ and $\log g$ (see Fig.~\ref{fig:models}).  Under these assumptions, our analysis methodology focusses on finding the $T_{\mathrm{eff}}$--$\log g$ pair for which the theoretical model best reproduces the diagnostic features. 
The synthetic spectra were convolved adopting a Gaussian
instrumental profile corresponding to the variant MUSE spectral
resolution over the wavelength range of the data.

%\textbf{Lastly, our models do not account for the effects of the stellar winds. However, at these low luminosities we will not see any effect of the winds in the spectral lines \citep{1982ApJ...253L..39S}.}

\subsection{Effective temperature and surface gravity determination}
\begin{figure}[t]
\centering
\includegraphics[width=.95\linewidth]{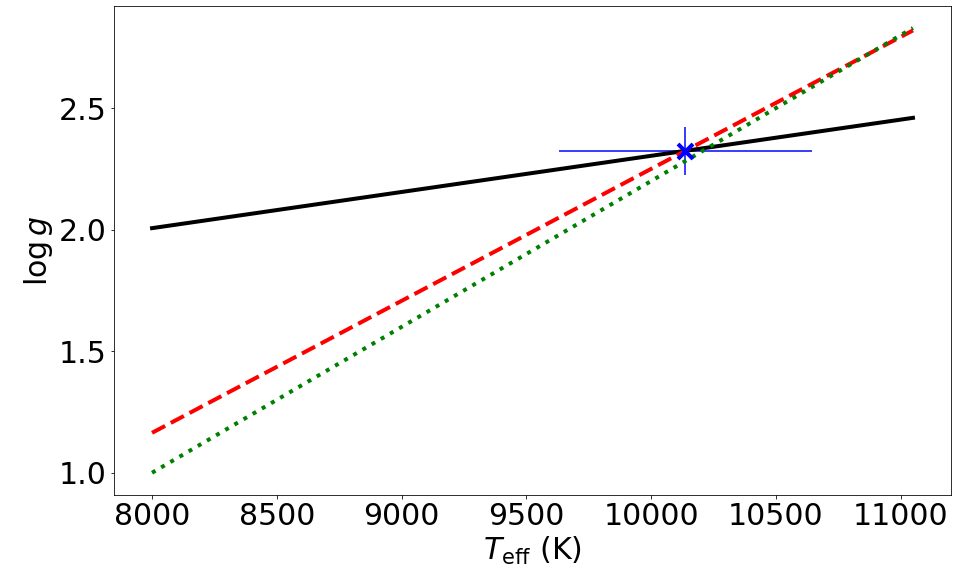}
\caption{Example of the method used to determine the temperature and surface gravity for star \#151. The locus obtained by varying $\log g$ (in cgs units) for each fixed model-grid $T_\mathrm{eff}$ in order to fit the hydrogen lines is shown by the solid black line. The loci obtained by varying $T_\mathrm{eff}$ at each fixed model-grid $\log g$ in order to fit the He\,{\sc i} and O\,{\sc i} lines, and to fit the metal-line dominated  4950-5600\,{\AA} region, are indicated by the red-dashed and the green-dotted lines, respectively. The blue cross marks the intersection of the loci, yielding the adopted atmospheric parameter values.
\label{fig:id151}}
\end{figure}

 We followed a well-established methodology to find the best solution for each object. %in the $T_{\mathrm{eff}}$--$\log\,\mathrm{g}$ plane. 
 First, by adopting $T_{\mathrm{eff}}$, we found the model that best reproduces the gravity sensitive features 
 (the hydrogen Balmer lines; this step was repeated for different adopted $T_{\mathrm{eff}}$ values, hence allowing us to define the locus of models for which the Balmer lines are equally well represented. We refer to this as the $\log g$ locus. In a similar fashion, but adopting the surface gravity, we identified the model that best reproduces the $T_{\mathrm{eff}}$ sensitive features (see below). Repeating this step for different values of $\log g$, we defined the $T_{\mathrm{eff}}$ locus. Finally, the intersection of both lines defines the best possible solution for a given object in the  $T_{\mathrm{eff}}$--$\log g$ plane (see Fig.\,\ref{fig:id151}).
 
\begin{figure}[t]
\centering
\includegraphics[width=.96\linewidth]{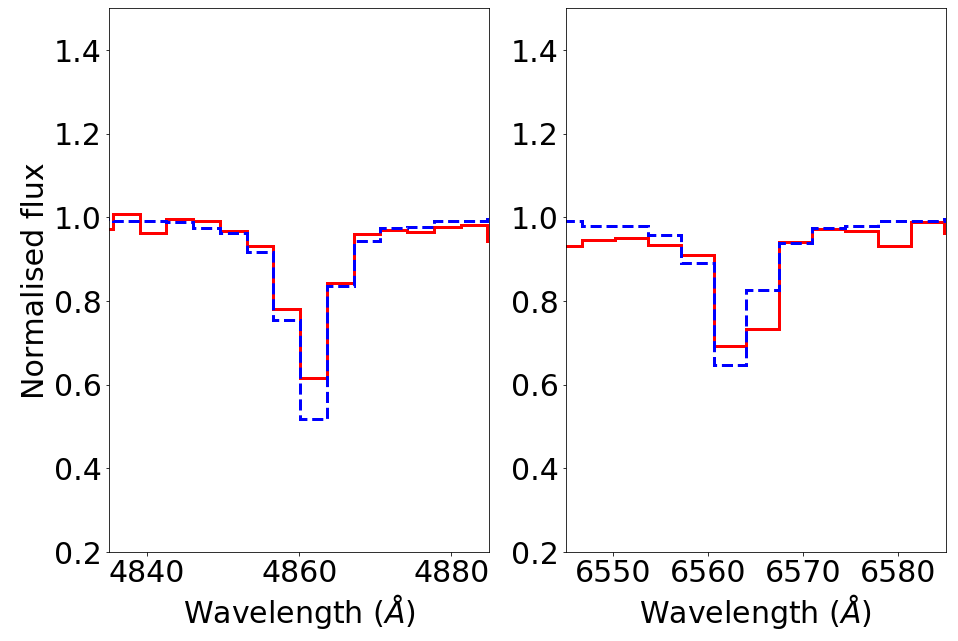}
\caption{Example for a best model fit to observed hydrogen lines for star \#192. The observed spectrum is marked by the solid red line, and the model is shown by the blue-dashed line. }
\label{fig:hlines1}
\end{figure}

\begin{figure}[t]
\centering
\includegraphics[width=.96\linewidth]{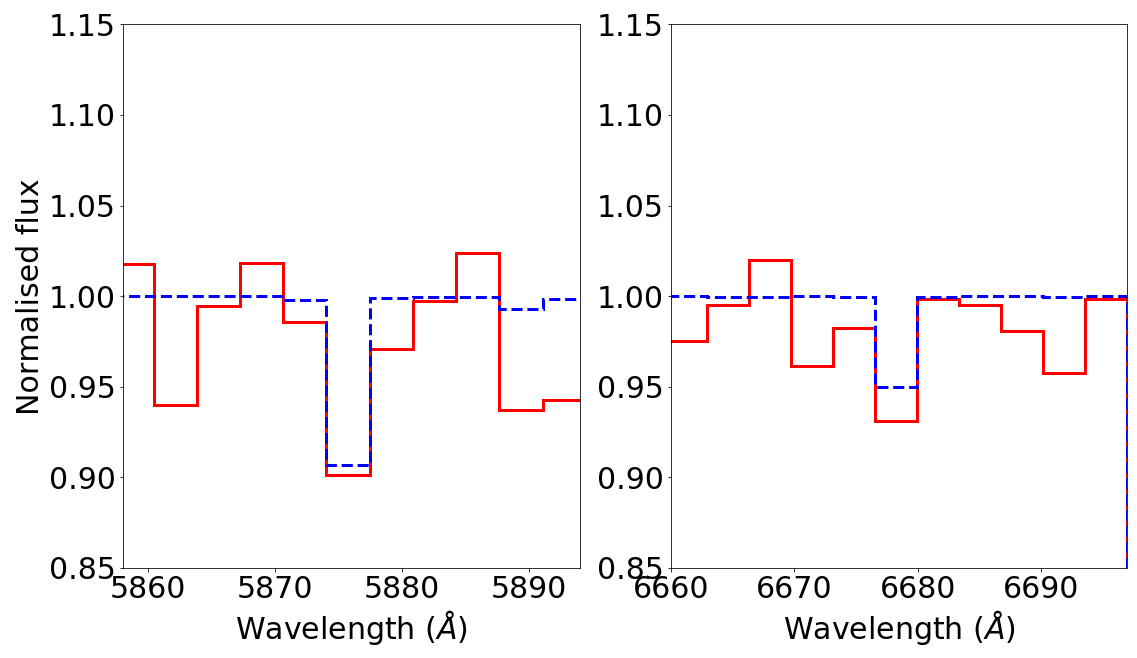}
\caption{Same as Fig.~\ref{fig:hlines1}, but for He\,{\sc i} lines.}
\label{fig:helines1}
\end{figure}

\begin{figure}[t]
\centering
\includegraphics[width=.96\linewidth]{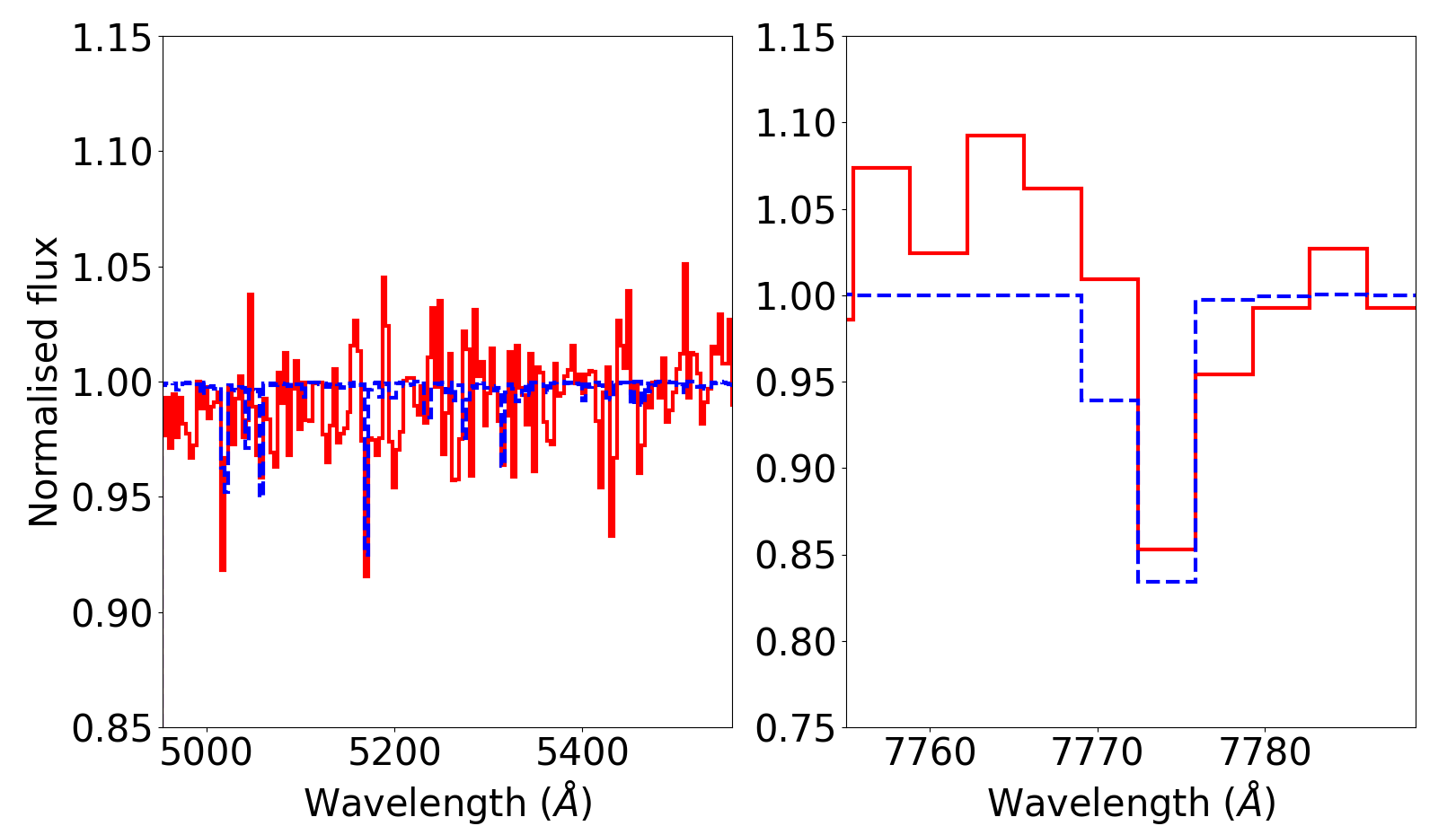}
\caption{Same as Fig.~\ref{fig:hlines1}, but for the region from 4950-5600\,{\AA} (left panel), and for O\,{\sc i} (right panel).}
\label{fig:siiilines1}
\end{figure}

 At each step, we focussed on fitting individual lines. When fixing each effective temperature in 
 our model grid to find the best-fitting surface gravity by means of a $\chi^{2}$ minimisation, we minimised 
 the residuals between observed and synthetic Balmer lines H$\alpha$ and 
 H$\beta$\footnote{ Unlike in previous applications of the models, our 
 MUSE data only cover H$\alpha$ and H$\beta$, which may be influenced by a sufficiently strong stellar wind.
 However, at the comparatively low luminosities of the target stars,
 the two lines are expected to be symmetric, that is~overall unaffected by mass outflow, 
 as a comparison with high-resolution spectra of analogous Galactic objects
 shows \citep{Verdugoetal99}. Consequently, they are represented well by
 the hydrostatic approach and the synthetic non-LTE profiles based on 
 the hydrogen model atom by \citet{PrBu04}.}. 
 Given the known correlation between the effective temperature and surface gravity, a linear relation with a positive slope is expected: for a higher $T_{\mathrm{eff}}$, we would need a higher $\log g$ to obtain the same strength of the lines.
 When finding $T_{\mathrm{eff}}$ for an adopted $\log g$ value, we consider mostly metal lines as sensitive features. The strongest lines observed in our spectra correspond to He\,{\sc i} and O\,{\sc i}, for which the synthetic spectra are based on the model atoms of \citet{Przybilla05} and \citet{Przybillaetal01}, respectively. A linear relation between $T_{\mathrm{eff}}$ and $\log g$ is again expected, thus the higher the assumed surface gravity, the higher the required effective temperature to reproduce the selected features. However, we did not find this linear relation for lower S/N spectra, with increasing noise giving rise to larger uncertainties. In addition to the lower S/N spectra, for decreasing $T_{\mathrm{eff}}$, the He\,{\sc i} lines become weaker and eventually impossible to detect. For this reason, we also considered fitting the lines in the spectral region from 4950\,\AA{} to 5600\,\AA, since numerous metal lines are present. We fitted the best continuum between the model and data by means of a $\chi^{2}$ minimisation, and the strongest metal lines in the region. In the cases in which it was possible to do so (high S/N targets), for the sake of completeness, we checked that both fitting strategies provide the same stellar parameters within the error limits (as seen in Fig.~\ref{fig:id151}).
Examples of the fit provided by the best model are shown in Figures~\ref{fig:hlines1}, \ref{fig:helines1}, and \ref{fig:siiilines1} for H, He\,{\sc i}, O\,{\sc i}, and the 
4950--5600\,\AA~spectral window. Table~\ref{tab:stellar-parameters} summarises the parameters derived in our analysis. %wavelengths Halpha 6562.8 \AA{}, Hbeta 4861.3 \AA{}, He\,{\sc i} 5875.6, 6678.0 \AA{}, O\,{\sc i} 7771.9 \AA{}.
We adopted a systematic error in $T_{\mathrm{eff}}$ of 5\%, and for $\log g$ (cgs) 0.10\,dex, both reasonable estimations given the grid of the models and the quality of the data, see for example \citet{Kudritzki08}.

%An initial sample of 26 of the brightest stars were considered for the analysis, all previously classified as B/A-type supergiants or bright giants by \citet{2018A&A...618A...3R}. Among these 26 targets, for four of them we obtain an extrapolated $T_{\mathrm{eff}}$ that is outside of the limits of our grid of models (objects id 252, 533, 1262 and 1526 in Fig.~\ref{fig:photometry}). We exclude these cases from further consideration since it cannot be assumed that the behaviour will be linear when extrapolated to higher/lower $T_{\mathrm{eff}}$.

%\begin{figure}[t]
%\centering
%\includegraphics[width=.9\linewidth]{figures_lines/cont_id_192.png}
%\caption{Example case of the continuum region 4950-5600\,\AA{} fitted to the best model. }
%\label{fig:cont1}
%\end{figure}

%-------------------------------------------------------------------------------

\begin{table*}[t]
\caption{Atmospheric and stellar fundamental parameters for the sample
stars.}
        \label{tab:stellar-parameters}
        \small
        \centering
        \begin{tabular}{l r r r r r r r r r l} 
                \hline \hline
                ID & $T_\mathrm{eff}$\,(K) & $\log\,g$ & $\log\,g_\mathrm{F}$ & $E(B-V)$ & $B.C.$ & $M_{\mathrm{bol}}$ & $\log\,L/L_{\odot}$ & $R/R_{\odot}$ & $M_\mathrm{spec}/M_{\odot}$ & Sp.~Type \\
\hline
%\hline
119 & 8990$\pm$450 & 2.16$\pm$0.10 & 2.34$\pm$0.13 & 0.14$\pm$0.03 & $-$0.08 & $-$6.37$\pm$0.22 & 4.45$\pm$0.09 & 69$\pm$7 & 25.1$\pm$3.3 & A2\,Ib~~~~~~~*\\
%\hline
151 & 10140$\pm$510 & 2.32$\pm$0.10 & 2.30$\pm$0.13 & 0.11$\pm$0.03 & $-$0.30 & $-$6.15$\pm$0.23 &  4.35$\pm$0.09 & 49$\pm$5 & 18.2$\pm$2.5 & A0\,Ib~~~~~~~*\\
%\hlineFamilien
192 & 10980$\pm$550 & 2.23$\pm$0.10 & 2.07$\pm$0.13 & 0.12$\pm$0.03 & $-$0.46 & $-$6.30$\pm$0.23 &  4.42$\pm$0.09 & 45$\pm$5 & 12.4$\pm$1.7 & B9\,Ib\\
%\hline
209 & 11840$\pm$590 & 2.55$\pm$0.10 & 2.26$\pm$0.13 & 0.25$\pm$0.03 & $-$0.63 & $-$6.84$\pm$0.24 &  4.63$\pm$0.10 & 50$\pm$5 & 31.3$\pm$4.3 & B9\,Ib~~~~~~~*\\
%\hline
223 & 12500$\pm$620 & 2.66$\pm$0.10 & 2.27$\pm$0.13 & 0.09$\pm$0.03 & $-$0.75 & $-$6.31$\pm$0.24 &  4.42$\pm$0.10 & 35$\pm$4 & 20.1$\pm$2.9 & B8\,Ib~~~~~~~*\\
%\hline
247 & 10370$\pm$520 & 2.23$\pm$0.10 & 2.17$\pm$0.13 & 0.09$\pm$0.03 & $-$0.34 & $-$5.86$\pm$0.23 &  4.25$\pm$0.09 & 41$\pm$4 & 10.6$\pm$1.5 & A0\,Ib\\
%\hline
%352 & 8360$\pm$420 & 1.80$\pm$0.10 & 2.11$\pm$0.13 & $-$0.02$\pm$0.03 & 0.01 & $-$4.97$\pm$0.22 &  3.88$\pm$0.09 & 42$\pm$4 & 4.0$\pm$0.5 & A3\,II\\
%\hline
%435 & 9830$\pm$490 & 1.86$\pm$0.10 & 1.89$\pm$0.13 & 0.32$\pm$0.03 & 0.27 & $-$5.94$\pm$0.23 &  4.27$\pm$0.09 & 47$\pm$5 & 5.9$\pm$0.8 & A1\,Ib\\
%\hline
462 & 8360$\pm$420 & 1.80$\pm$0.10 & 2.11$\pm$0.13 & 0.51$\pm$0.03 & 0.02 & $-$6.30$\pm$0.22 &  4.42$\pm$0.09 & 77$\pm$8 & 13.6$\pm$1.8 & A3\,Ib\\
%\hline
468 & 12500$\pm$630 & 2.66$\pm$0.10 & 2.27$\pm$0.13 & 0.19$\pm$0.03 & $-$0.75 & $-$6.11$\pm$0.24 &  4.34$\pm$0.10 & 32$\pm$4 & 16.6$\pm$2.3 & B8\,Ib~~~~~~~*\\
%\hline
507 & 9080$\pm$450 & 2.17$\pm$0.10 & 2.34$\pm$0.13 & 0.09$\pm$0.03 & $-$0.09 & $-$5.10$\pm$0.22 &  3.93$\pm$0.09 & 38$\pm$4 & 7.5$\pm$1.0 & A2\,II\\
%\hline
629 & 8730$\pm$440 & 2.12$\pm$0.10 & 2.36$\pm$0.13 & 0.10$\pm$0.03 & $-$0.03 & $-$4.82$\pm$0.22 &  3.83$\pm$0.09 & 36$\pm$4 & 6.2$\pm$0.8 & A3\,II\\
%\hline
633 & 9430$\pm$470 & 2.30$\pm$0.10 & 2.40$\pm$0.13 &  0.21$\pm$0.03 & $-$0.16 & $-$5.54$\pm$0.22 &  4.07$\pm$0.09 & 41$\pm$4 & 12.1$\pm$1.6 & A1\,II~~~~~~~*\\
%\hline
641 & 8920$\pm$450 & 2.14$\pm$0.10 & 2.34$\pm$0.13 & 0.15$\pm$0.03 & $-$0.06 & $-$5.10$\pm$0.22 &  3.94$\pm$0.09 & 39$\pm$4 & 7.6$\pm$1.0 & A3\,II\\
%\hline
645 & 9810$\pm$490 & 2.14$\pm$0.10 & 2.17$\pm$0.13 & 0.22$\pm$0.03 & $-$0.23 & $-$5.62$\pm$0.23 &  4.14$\pm$0.09 & 41$\pm$4 & 8.4$\pm$1.2 & A1\,II\\
%\hline
709 & 8730$\pm$440 & 2.11$\pm$0.10 & 2.35$\pm$0.13 & 0.09$\pm$0.03 & $-$0.03 & $-$4.95$\pm$0.22 & 3.88$\pm$0.09 & 38$\pm$4 & 6.7$\pm$0.9 & A3\,II\\
%\hline
759 & 8300$\pm$420 & 2.05$\pm$0.10 & 2.37$\pm$0.13 & 0.15$\pm$0.03 & 0.03 & $-$4.95$\pm$0.22 &  3.90$\pm$0.09 & 42$\pm$4 & 7.2$\pm$1.0 & A4\,II\\
%\hline
200230 & 8300$\pm$420 & 2.05$\pm$0.10 & 2.37$\pm$0.13 & 0.41$\pm$0.03 & 0.03 & $-$6.03$\pm$0.22 &  4.31$\pm$0.09 & 70$\pm$7 & 19.4$\pm$2.6 & A4\,Ib~~~~~~~*\\
%\hline
%200272 & 9310$\pm$470 & 1.83$\pm$0.10 & 1.95$\pm$0.13 & 0.02$\pm$0.03 & 0.18 & $-$4.95$\pm$0.22 &  3.88$\pm$0.09 & 33$\pm$3 & 2.7$\pm$0.4 & A1\,II\\
%\hline
%200275 & 10360$\pm$520 & 1.44$\pm$0.10 & 1.38$\pm$0.13 &  0.10$\pm$0.03 & 0.40 & $-$5.60$\pm$0.23 &  4.14$\pm$0.09 & 36$\pm$4 & 1.3$\pm$0.2 & A0\,II\\
\hline
        \end{tabular}
\tablefoot{ID numbers are taken from Paper~I. Spectral types were assigned based on the empirical $T_\mathrm{eff}$ spectral type and $\log L/L_\odot$ luminosity class relations (see text). An asterisk marks cases with a large mismatch of spectroscopic end evolutionary masses (objects marked in red in Figs.~\ref{fig:hr} to \ref{fig:myfglr}). The results for the corresponding entries should be considered with care; the objects are likely not single stars as discussed in Sect.~\ref{sec:discussion}.}
\end{table*}

\subsection{Stellar luminosities, radii, and masses}
With the fundamental stellar parameters $T_\mathrm{eff}$ and $\log g$ at hand, we proceeded to  derive luminosities $L$, spectroscopic masses $M_\mathrm{spec}$, and radii $R$ for the stars in our sample by combining observed magnitudes in different photometric bands with theoretical quantities obtained from the spectral energy distribution corresponding to the best-fitting models.
A catalogue of observed Johnson $B$- and $V$-band magnitudes for sources in NGC\,300, based on the work by \citet{Piet01}, was kindly provided by F.~Bresolin (private communication). The coordinates of the sources in this catalogue were cross-matched with our MUSE observations in order to extract the magnitudes for our targets. Coordinates and photometric magnitudes are reported in Table~\ref{tab:photometry}. %It was not possible to cross-identify two of the stars in our original sample (ids 830 and 1154), hence we decided to drop them from the further discussion.

Individual reddening values $E(B-V)$ were obtained by comparing observed and theoretical $(B-V)$ colours from tailored models computed for the adopted atmospheric parameters. Bolometric corrections $B.C.$ were also calculated from the tailored models. We adopted a standard value for the ratio of selective-to-total extinction $R_{V}$\,=\,3.1 to calculate the $V$-band extinction, and a tip of the red-giant branch (TRGB) distance to NGC\,300 of $d$\,=\,1.86$\pm$0.07\,Mpc as determined by \cite{Rizzi2006}. Bolometric magnitudes $M_\mathrm{bol}$ were calculated from the extinction-corrected apparent $V$-band magnitude, the distance, and the $B.C.$ for each object individually. The bolometric magnitudes were consecutively converted to stellar luminosities. Finally, from the combination of the stellar luminosities with the derived atmospheric parameters $T_\mathrm{eff}$ and $\log g$, stellar radii and spectroscopic masses were calculated. 

Table~\ref{tab:stellar-parameters} summarises all the derived properties of the stars in our sample. The table also presents the flux-weighted gravity of each star, $g_{\mathrm{F}}$\,=\,$g\times\left(T_{\mathrm{eff}}/10^{4}\right)^{-4}$, introduced by \citet{2003ApJ...582L..83K}. We note that this quantity is a proxy for luminosity, as discussed in the aforementioned reference, and when normalised to the solar value, it represents the inverse of the quantity used by \cite{Langer14} to define the spectroscopic Hertzsprung--Russell diagram (sHRD). The uncertainties presented in the table were calculated using Gaussian error propagation. For the observed magnitudes, we adopted the errors provided by \citet{Piet01}; the uncertainties affecting the intrinsic colours and bolometric corrections have a negligible effect on the final uncertainties.   

Regarding the reddening values presented in Table~\ref{tab:stellar-parameters}, our stars show a wide range. However, disregarding the two stars with the highest values, the rest are consistent with the results presented by \citet{Kudritzki08} for BA-type supergiant stars in NGC\,300, as well as with the characteristic value for Cepheids based on optical and near-IR photometry by \citet{gieren2005}.

%-----------------------------------------------------------------------
\subsection{Spectral type and luminosity class determination}
The wavelength coverage of MUSE does not include the blue spectral region. Combined with the low S/N of the sample stars, a classical determination of spectral sub-types on a pure observational basis is not feasible here. Nevertheless, a model-based approach can be used;  
we applied the relationship between the spectral type and effective temperature as obtained by \citet{Firnstein12} based on Galactic BA supergiants. We have a few cases for which the derived effective temperatures are lower than the lowest average value for the coolest type in \citet{Firnstein12}, A3. For these cases, we adopted either an A3 or A4 spectral type. 
Assigning a luminosity class LC to each object is also challenging. We resorted to the work by \citet{Firnstein10}, which gives guidance to constrain the luminosity class from the stellar luminosity. Cases of LC II result from a reasonable extrapolation. The results are summarised in Table~\ref{tab:stellar-parameters}.

%-------------------------------------------------------------------

\section{Discussion}\label{sec:discussion}
\subsection{Comparison with previous studies}
A previous study of the same sample was presented by \citet{2018A&A...618A...3R}. These authors analysed the stars by means of two different spectral libraries: (1) the empirical library MIUSCAT (unpublished, provided by Alexandre Vazdekis), and (2) a grid of PHOENIX models \citep{Husser13}. 

MIUSCAT is an extension of the MILES library (\citet{Sanchez06,Cenarro07,Falcon11}). The library of PHOENIX models covers a range of 2300\,K\,$\leq$\,$T_{\mathrm{eff}}$\,$\leq$\,12000\,K and 0.0\,$\leq$\,$\log g$\,$\leq$\,6.0 using the assumption of LTE.
In both cases, the UlySS code \citep{Koleva09} based on pPXF \citep{2004PASP..116..138C}, initially developed for the study of stellar clusters, was used to find the best matching solution. Unlike the work presented here, \citet{2018A&A...618A...3R} proceeded to fit the full un-normalised  spectra. In the case of hot massive stars, the main concern in terms of inaccuracy of their results would be the assumption of LTE.

Comparing our results with those by \citet{2018A&A...618A...3R}, $T_{\mathrm{eff}}$ values derived in both studies are consistent within the uncertainties for only eight stars (50$\%$ of our final sample), whilst in terms of $\log g$ this occurs in only four cases (25$\%$). The tendency is for us to obtain higher gravities. These differences can be understood in terms of the very different methodology employed to analyse the stars, as well as the differences in the physical assumptions of the models. In addition, the MIUSCAT library has incomplete coverage of the HRD for hot stars \citep[see discussion by][]{2018A&A...618A...3R}.
%It could be that the continuum is over-estimated for the normalized spectra.

\begin{figure}[t]
\centering
\includegraphics[width=0.98\linewidth]{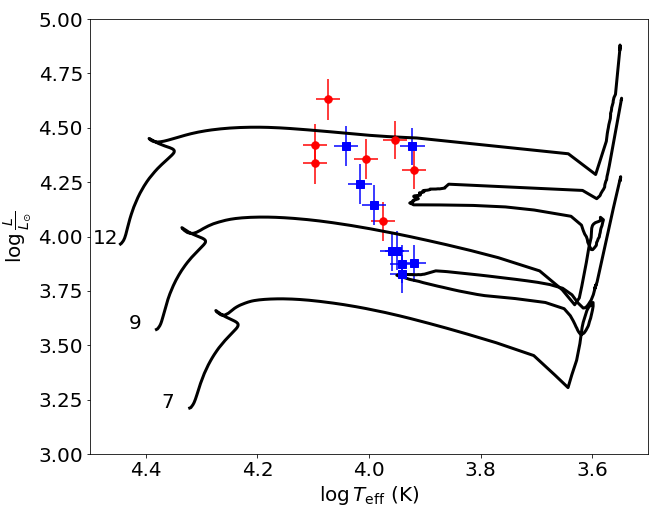}
\caption{HRD with evolutionary tracks for different masses (black solid lines, in $M_\odot$) that account for rotation \citep{Evoltracks12}. The red dots represent sample stars that move by more than 10$\%$ their evolutionary mass track position with respect to the sHRD in Fig.~\ref{fig:shr}, and the blue squares are those that move less than 10$\%$.}
\label{fig:hr}
\end{figure}

Regarding the spectral classification, we found four late B-type supergiants, ten early A-type stars (four of them supergiants and the rest bright giants), and two A4-type stars (with one supergiant and one bright giant). Comparing this with the previous classification, we resolved the A-types to a finer sub-class. On the other hand, \citet{2018A&A...618A...3R} found a broader range of luminosity classes, varying from II to Ia. This could be attributed to the very limited suitability of the empirical libraries for hot stars as pointed out above. The fact that the classification has been derived directly by model fitting while in our case we used the luminosity values of known stars as a reference could also explain our broader luminosity class range. We note, however, that the brightest supergiants in NGC\,300, that is to say likely those of LC Ia, are of 18$^\mathrm{th}$\,mag \citep{2002ApJ...567..277B},
so 2 to 4\,mag brighter than the stars investigated here.

%---------------------------------------------------

%To wrap up, we have derived the stellar parameters such as effective temperature, surface gravity, luminosity, radius, masses and bolometric magnitudes of the 20 brightest stars of NGC 300 field-i MUSE data-cube. We have been able to classify those stars as B and A-type supergiants and bright giants and have compared the results with the previous work done by \cite{2018A&A...618A...3R}, concluding that our stellar parameter determination method is more accurate than the previous because we have checked individual normalized lines to reduce as much as possible the noise effects.
 
%The next section is going to show an application to the analysis done. To understand this application we need to first introduce an important concept called the Flux-Weighted Gravity Luminosity Relationship (FGLR). 

\subsection{Extending the flux-weighted gravity luminosity relationship (FGLR)}
The FGLR was first derived by \citet{2003ApJ...582L..83K} as a new method for distance determination of supergiants:
\begin{equation}
\label{eq:fglrfinal}
-M_{\mathrm{bol}}=a(\log g_{\mathrm{F}}-1.5)+b
\end{equation}
with $a$ determined by the mass-luminosity relation ($L$\,$\propto$\,$M^x$) exponent $x$ to $a$\,=\,2.5$x/(1-x)$. For massive stars, $x$\,$\approx$\,3 is found.
%\begin{equation}
%\label{eq:a}
%a=2.5x/(1-x).
%\end{equation}

This relation holds for all supergiants and bright giants that have a constant luminosity track when they move to the right of the HRD. We can then use this relation to estimate the bolometric magnitudes, luminosities, and distances of supergiants and bright giants from which we only have spectral information. If we are able to resolve massive stars in distant galaxies, this can be a very powerful spectroscopic method to determine extragalactic distances.

\begin{figure}[t]
\centering
\includegraphics[width=0.975\linewidth]{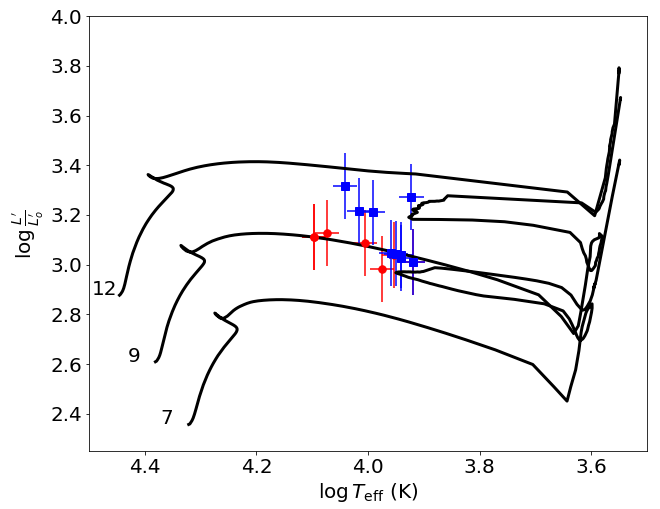}
\caption{Same as Fig.~\ref{fig:hr}, but for the sHRD. The $L'$ is defined as the inverse of the flux-weighted gravity. The symbol encoding refers to movements relative to the tracks with respect to Fig.~\ref{fig:hr}.}
\label{fig:shr}
\end{figure}

To study the FGLR, we need to be sure that the stars are at the correct evolutionary stage. To verify that, we plotted the stars in relation to evolutionary tracks from \citet{Evoltracks12} in the regular HRD (Fig.~\ref{fig:hr}) and compared their position with respect to the tracks in the sHRD (Fig.~\ref{fig:shr}). 

The sHRD \citep{Langer14} shows the inverse of $g_{\mathrm{F}}$ with respect to $T_\mathrm{eff}$. Using the sHRD, we can place the stars with only their spectroscopic information and without any knowledge of their distance or brightness. This provides an advantage to the HRD where we need to assume a distance to obtain the luminosities. In the HRD, if two stars with same $T_{\mathrm{eff}}$ and $L$, but different masses (e.g. supergiants and post-AGB stars around $\log L/L_\odot$\,$\sim$\,4), occupy the same location, they must have the same radius, so the $\log g$ has to be different. In the sHRD, the same degeneracy does not occur since the stars fall onto different iso-gravity lines, enabling us to discriminate them. Also multiple systems, where one star dominates the observed spectrum but other objects contribute to the total luminosity, can be discriminated.

We have defined a threshold of 10$\%$ of the star mass for the change in their relative position with respect to their evolutionary tracks in both diagrams, which is reasonable since it corresponds to their mass error. The red dots in Figs.~\ref{fig:hr} and \ref{fig:shr} represent stars that move more than the 10\% threshold, which indicates that they are not well-behaved objects. These are the same stars that show significantly larger spectroscopic than evolutionary masses that are derived from comparison with evolutionary tracks.
The blue squares in Figs.~\ref{fig:hr} and \ref{fig:shr} do not move by more than the threshold and a good correspondence between their spectral information and their true evolutionary stage and their single star status can be assumed. The latter stars are certain to be in the supergiant stage and hence the FGLR would hold. 

\begin{figure}[t]
\centering
\includegraphics[width=.90\linewidth]{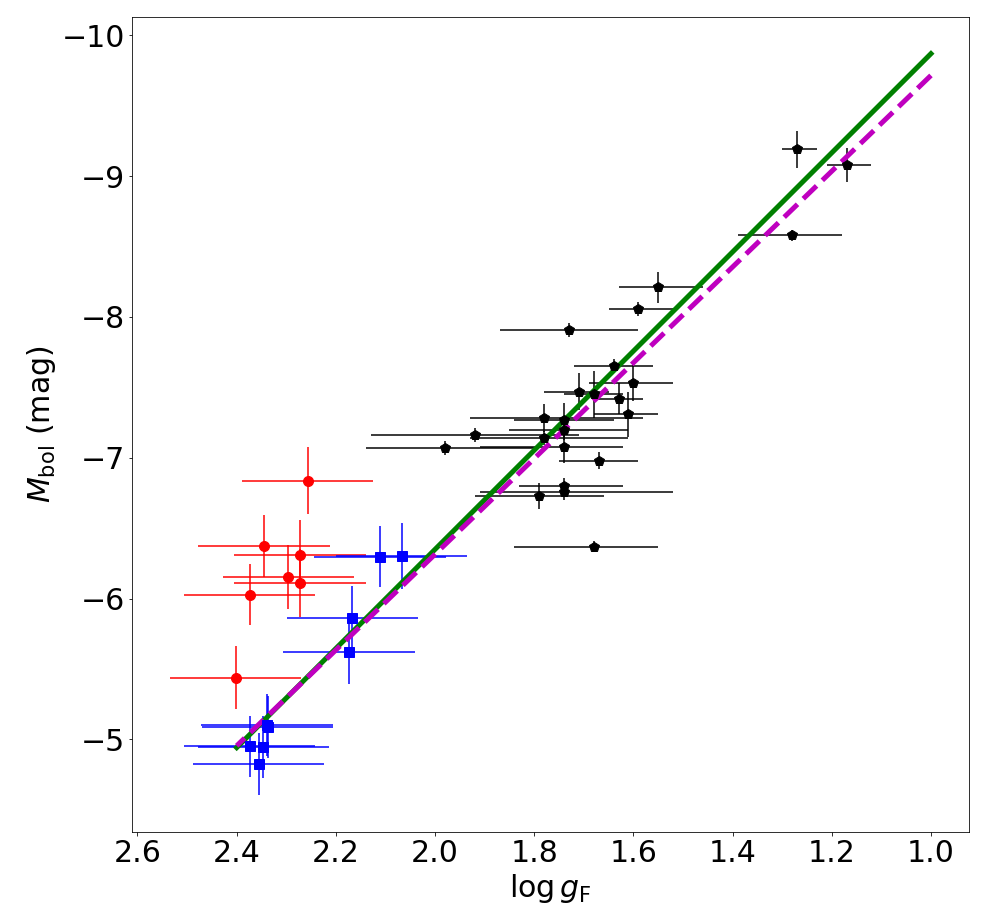}
\caption{FGLR for the stars in NGC\,300. Black symbols denote the stars studied by \citet{Kudritzki08}, with the corresponding FGLR regression line marked in solid green.
Blue and red symbols mark the stars analysed here, with the magenta-dashed line representing the regression line derived from the black and blue symbols.}
\label{fig:myfglr}
\end{figure}

To prove our last point, we considered our 16 stars along with the
objects previously studied by \citet{Kudritzki08} to derive the FLGR.
As we can see in Fig.~\ref{fig:myfglr}, the blue squares follow the
old FGLR within their error limits. The initial FGLR gives the old parameters determined by \citet{Kudritzki08}: $a_{\mathrm{old}}$\,=\,$-3.52$ and $b_{\mathrm{old}}$\,=\,8.11. Adding the contribution of our newly found supergiants (blue squares in Fig.~\ref{fig:myfglr}), we obtained $a$\,=\,$-$3.40$\pm$0.04 and $b$\,=\,8.02$\pm$0.14. The results from this work are also in accordance with the new FGLR distance to NGC\,300 $(m-M)_{\mathrm{FGLR}}$\,=\,26.34$\pm$0.06 by \citet{2021ApJ...914...94S}.

As previously discussed, we performed the analysis with $[Z]$\,=\,0.0\,dex and $[Z]$\,=\,$-$0.15\,dex, and although we found small systematic differences between the results, there is no observable effect on the derived flux-weighed gravities. This is due to the fact that $T_{\mathrm{eff}}$ and $\log g$ are covariant; an increase in $T_{\mathrm{eff}}$ requires an increase in $\log g$, and vice versa. Hence, $\log g_{\mathrm{F}}$ does not effectively change (in particular given the uncertainties) and nor does the FGLR since the $T_{\mathrm{eff}}$ changes are too small to introduce any effect on the derived reddening values.

\subsection{Discrepant cases}
As seen in Fig.~\ref{fig:myfglr}, stars depicted by blue squares follow -- within the uncertainties -- the trend defined by the FGLR, whilst the ones represented by red dots deviate to some extent.
We shall discuss these discrepant cases. For a star to lie above the relationship, under the consideration that it should not be there, either the derived luminosity is too high, or the flux-weighted gravity is too large (or both). The primary reason for the luminosity (bolometric magnitude) to be too high would be for the apparent magnitude to be too high. That could result from the fact that what is seen as a single star is in fact a combination of several unresolved sources. In that case, it could well happen that whilst the most luminous star dominates the observed spectrum, one or several fainter sources could be contributing to the continuum, and hence the observed apparent magnitudes. Inspection of the MUSE datacube (see Fig.~\ref{fig:photometry}) results in that none of these objects show signs of being an extended source, which makes it unlikely that they are large star clusters. On the other hand, it cannot be ruled out that they are small stellar aggregates that are not resolvable at the distance of NGC\,300. 

An alternative explanation for this deviation is that as we increase the $\log g_{\mathrm{F}}$ and move to the bottom left of the FGLR, the stars decrease in mass. Population simulations predict that the FGLR will get wider for lower masses, as discussed by \citet{Meynet15}. We expect the density of objects to increase as the masses decrease because of the so-called initial mass function (IMF) effect: we always expect to find a higher number of low mass stars than of massive stars \citep[e.g.][]{1955ApJ...121..161S,2001MNRAS.322..231K}, widening the FGLR because of the increased scatter.
%Based on this effect we could justify increased spread in our stars in Fig.~\ref{fig:myfglr}, being able to explain all the deviating red cases due to a wider FGLR at low masses. 1

%\begin{figure}[h]
%\centering
%\includegraphics[width=1.\linewidth]{project/figs/meynet-15-fig8-2.png}
%\caption{Simulated populated synthesis of blue supergiants in the FGLR compared to observations (black triangles) and including the effects of observational errors. Left panel: non-rotating models for $Z=0.014$. Right panel: rotating models for $Z=0.014$. Credit: \cite{Meynet15}.}
%\label{fig:sim}
%\end{figure}
%----------------------------------------------------------------------
\section{Outlook}
\label{sec:conclusion}
A quantitative analysis of 16 BA-type supergiants and bright giants in NGC\,300 based on VLT/MUSE integral field spectroscopy was performed here, with a focus on determining basic atmospheric and fundamental stellar parameters. This allowed us to verify that the FGLR can be extended towards less luminous stars than studied before. However, the study has faced limitations by the rather restricted $S/N$\,$\lesssim$\,20 of the spectra. 
For future work, longer exposures and use of the currently available AO mode of MUSE
should be aimed at boosting the $S/N$ and improving the spatial resolution. This would not only facilitate a similar study as the one performed here to be achieved at much reduced uncertainties,
but this would also help to determine metallicities and likely elemental abundances for selected individual chemical elements. The full potential of MUSE for extragalactic stellar astrophysics could thus be demonstrated. Studies with increased scientific value could be made once
the BlueMUSE medium-resolution panoramic integral field spectrograph becomes operational, which has been proposed as a new instrument for the VLT \citep{bluemuse}. BlueMUSE is optimised for the optical blue, where most of the diagnostic spectral features of hot supergiants are located.

\begin{acknowledgements}
We would like to thank the referee for useful comments which
helped improve the paper. Based on observations collected at the European Southern Observatory under ESO programme 094.D-0116(A).
The authors want to thank Tim-Oliver Husser, Benjamin Giesers and
Fabio Bresolin for valuable input.
GGT acknowledges funding by a scholarship within the Erasmus Mundus
Joint Master Degree programme AstroMundus. SK acknowledges funding from UKRI in the form of a Future Leaders Fellowship (grant no. MR/T022868/1). NC acknowledges funding from the Deutsche Forschungsgemeinschaft (DFG) - CA 2551/1-1.
\end{acknowledgements}

\bibliographystyle{aa}
\bibliography{sample2.bib}

%\begin{appendix} %First appendix

%\section{Title of Second appendix.....} %Second appendix

%\end{appendix}
%
%
\end{document}